# Asymmetric nonlinear-mode-conversion in an optical waveguide with PT symmetry

Changdong Chen[1,2,*], Youwen Liu[1,2], Lina Zhao[3], Xiaopeng Hu[4], and Yangyang Fu[1,2,*]

1 College of Physics, Nanjing University of Aeronautics and Astronautics, Nanjing 211100, People's Republic of China

2 Key Laboratory of Aerospace Information Materials and Physics (NUAA), MIIT, Nanjing 211106, China

3 Collaborative Innovation Center of Light Manipulations and Applications, Shandong Normal University, Jinan 250358, People's Republic of China

4 National Laboratory of Solid State Microstructures and Department of Physics, Nanjing University, Nanjing, 210093, People's Republic of China

*Corresponding author: cdchen@nuaa.edu.cn, yyfu@nuaa.edu.cn

## Abstract

Asymmetric mode transformation in waveguide is of great significance for on-chip integrated devices with one-way effect, while it is challenging to achieve asymmetric nonlinear-mode-conversion (NMC) due to the limitations imposed by phase-matching. In this work, we theoretically proposed a new scheme for realizing asymmetric NMC by combining frequency-doubling process and periodic PT symmetric modulation in an optical waveguide. By engineering the one-way momentum from PT symmetric modulation, we have demonstrated the unidirectional conversion from pump to second harmonic with desired guided modes. Our findings offer new opportunities for manipulating nonlinear optical fields with PT symmetry, which could further boost more exploration on on-chip nonlinear devices assisted by non-Hermitian optics.

## 1. Introduction

Owing to the abundance of non-conservative processes in optical fields, the concept of parity-time (PT) symmetry, first proposed in quantum mechanics [1, 2], has been introduced into photonics and attracted considerable attentions in recent years [3, 4]. By implementing optical PT symmetry with refractive index profile $n(y) = n^*(-y)$, varieties of intriguing phenomena have been discovered in various optical systems [5-18], such as optical waveguides [5-7], lasers [8-10], optical resonators [11-14], dissimilar antenna-type resonators [15, 16]. The concept of PT symmetry is further developed as non-Hermitian optics [19], in which spatially engineering gain and/or loss are considered to manipulate optical fields. In particular, much attention has been taken in the non-Hermitian modulation with a periodical refractive index $\Delta n = \rho_1 \cos(qy) + i\rho_2 \sin(qy)$, which enables an one-way momentum at the exceptional point with $\rho_1 = \rho_2$. Utilizing this unique property in active/passive optical systems, many interesting wave phenomena and novel applications [19-31] have been investigated, for instance, non-reciprocal light propagation [20-24], unidirectional invisibility [25-27], asymmetric diffraction [28]. While all these prior arts developed from the one-way momentum are largely locked in linear waves, and it is significant to introduce them into nonlinear optics, which could offer new paradigms to explore new nonlinear effects, such as one-way second harmonic (SH) generation, mode-convertor and so on.

As known, the key issue of nonlinear interactions within waveguide (e.g. LiNbO$_3$ (LN) waveguide) is the momentum (phase) mismatch, resulting from dispersions both in materials and guided modes. The most common way to solve such a problem is employing poled quasi-phase-matching structures [32, 33], which can compensate for the momentum mismatch between fundamental wave (FW) and SH modes. Although some other ways [34, 35] have been proposed to realize the momentum match in the process of nonlinear interactions, all these structures therein are impossible to realize asymmetric performance for nonlinear-mode-conversion (NMC), i.e., FW to SH. Here we proposed and demonstrated a new way for achieving NMC with asymmetric response in a LN waveguide with PT symmetry. The underlying mechanism lies on that the frequency-doubling process in the waveguide is designed to obtain the higher-mode of SH from the fundamental mode of FW and then unidirectional mode conversion from the higher-mode to the fundamental mode of SH is generated by the one-way momentum from PT symmetry. This Letter opens up a new way to obtain SH generator and expands the current capabilities of nonlinear devices.

## 2. The model of asymmetric NMC

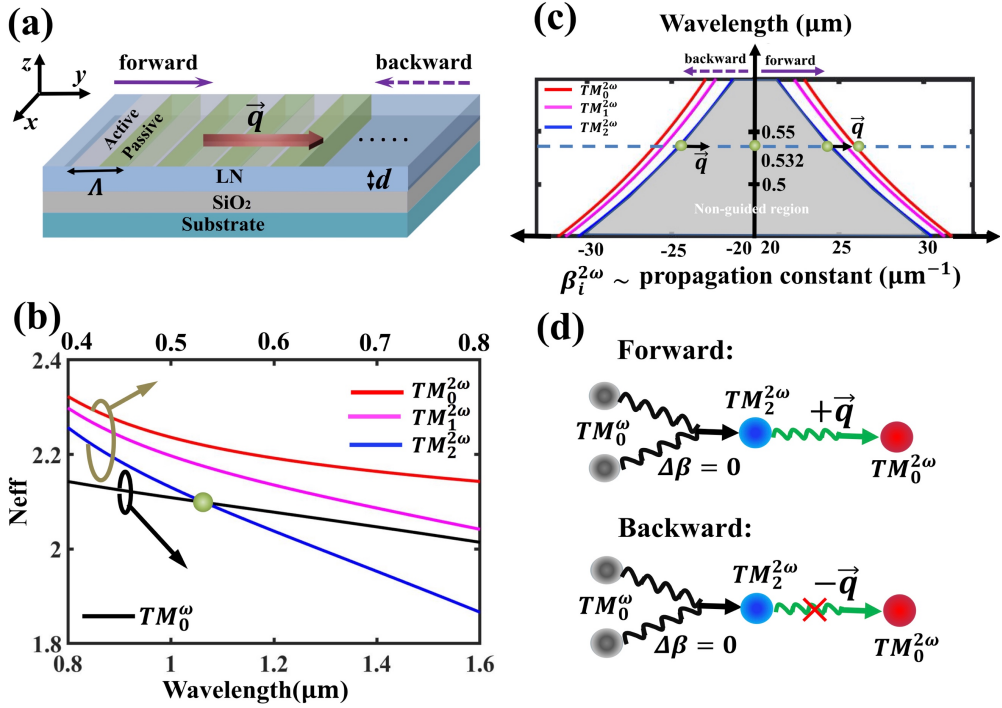

Fig. 1. (a) The schematic configuration of asymmetric NMC. Active and passive modulations taking a period of $\Lambda$, with $\Delta\varepsilon(y) = \cos(qy) + i\sin(qy)$, are applied on the LN waveguide layer with a thickness of $d$=1.01 $\mu m$, giving rise to a one-way wave-vector $\vec{q}$ along $y$ direction. (b) The mode dispersions in LN waveguide for FW (from 0.8 $\mu m$ to 1.6 $\mu m$) and the corresponding SH. There is only one intersection at about 1.064 $\mu m$ that is a probable FW wavelength to realize phase-matched frequency-doubling process. (c) Forward and backward phase-matching conditions induced by PT symmetry for $TM_2^{2\omega}$ mode in the dispersion diagram of SH. (d) The sketch map of entire process of the asymmetric NMC Only for forward-propagation, can two FW photons (*0$^{th}$* order mode) be transformed into one SH photon (*0$^{th}$* order mode) via the coupling between the interactions resulting from nonlinearity and PT symmetry.

Figure 1(a) shows schematically our considered structure, which is comprised of a single crystal *z*-cut LN film layer and a buried SiO$_2$ layer underneath LN film. Periodically arranged optical potentials $\Delta\varepsilon(y) = \cos(qy) + i\sin(qy)$, along the light propagation in the *y* direction, are implemented on the LN waveguide with dielectric constant $\varepsilon_r$. To be exact, the LN waveguide is in possession of a perturbative PT symmetry profile, i.e., $\varepsilon(y) = \varepsilon_r + \Delta\varepsilon(y)$. Such a periodically active and passive modulations can form a long-period grating with a period of $\Lambda$. This type of periodic modulation will possess a *y*-dependent perturbation function $\xi(y) = e^{iqy}$ where $q = 2\pi/\Lambda$, and here we set the origin as the starting point of the initial modulation period. Obviously from its Fourier transform, a unidirectional wave-vector (momentum) $\vec{q}$ can be introduced along y-direction, which can interact with the guided modes in LN film. Utilizing this feature, it is feasible to obtain unidirectional conversion from FW to SH with desired guided modes by devising specific nonlinear interactions and modulation periods in the waveguide.

Besides the one-way wave-vector, another critical issue is to realize pronounced frequency-doubling process for efficient nonlinear mode transformation. We show the mode dispersions of FW (wavelength regime from 0.8 to 1.6 *μm*) and SH (0.4 to 0.8 *μm*) of the LN waveguide in Fig. 1(b), where the waveguide parameters are 1.01 *μm* and 10 *μm* respectively for LN film and SiO$_2$ layer. Note that *TM* (electric field along *z*-polarization) waves are just considered here in order to exploit the largest nonlinear coefficient *d$_{33}$* of LN crystal. There is only one intersection point at FW of 1.064 *μm* (the green point in Fig. 1(b)) among distinct dispersion lines. This point is exactly the phase-matching point that allows to achieve considerable $TM_2^{2\omega}$ (2$^{nd}$ order of SH mode). Since the effective refractive index of FW mode remains lower than that of SH mode for the identical order, it is of great difficulty to realize efficient conversion of $TM_0^{\omega} \Rightarrow TM_0^{2\omega}$ directly, either by regulating waveguide parameters [34, 35], or assisted by meta-surface [36]. However in our scenario, if the generated SH wave satisfies $\overrightarrow{\beta_2^{2\omega}} + \vec{q} = \overrightarrow{\beta_0^{2\omega}}$, where $\overrightarrow{\beta_i^{2\omega}}(i = 0,1,2 ...)$ represents the propagation constant of the *i*$^{th}$ order of SH mode, then the $TM_2^{2\omega}$ mode of forward propagation can be converted into the $TM_0^{2\omega}$ mode (see the right part of Fig. 1(c)). While for backward propagation, as shown in the left part of Fig. 1(c), $\overrightarrow{\beta_2^{2\omega}}$ may be transmitted to the spatial frequency of $\overrightarrow{\beta_2^{2\omega}} - \vec{q}$, which is in the non-guided region of waveguide in our case. Accordingly, $\vec{q}$ has no influence on the $TM_2^{2\omega}$ mode of backward propagation.

The overall process of asymmetric NMC could be understood from in Fig. 1(d). Owing to phase matching $\Delta\beta = 0$, two FW photons of $TM_0^\omega$ mode are combined to generate directly one SH photon of $TM_2^{2\omega}$ mode, independent of the propagation direction. For the forward process, the SH photon of $TM_2^{2\omega}$ mode will be converted into $TM_0^{2\omega}$ mode via the additional wave-vector $\vec{q}$ from PT symmetry. While for the backward case, the wave-vector $\vec{q}$ cannot be involved in the interaction, the phase-matching is unable and then the $TM_2^{2\omega}$ mode of SH photon is unchanged.

## 3. Results and discussion

To well simulate the whole interactions, an approach analogous to the split-step Fourier method has been employed. In propagating over a small distance, the nonlinear frequency conversion and PT induced mode conversion can be assumed to act independently for the optical field, while they act together along the whole length of waveguide. Consequently, the dynamics of nonlinear interaction can be derived as:

$$\frac{dA_0^\omega}{dy} = \frac{i\omega}{Neff_0^{2\omega}c}\left[\kappa_1^0 A_0^{2\omega}(A_0^\omega)^* \cdot e^{-i\Delta\beta_0 y} + \kappa_1^2 A_2^{2\omega}(A_0^\omega)^* \cdot e^{-i\Delta\beta_2 y}\right] \quad (1)$$

$$\frac{dA_0^{2\omega}}{dy} = \frac{i\omega}{Neff_0^{2\omega}c}\left[\kappa_2^0 (A_0^\omega)^2 \cdot e^{i\Delta\beta_0 y}\right] \quad (2)$$

$$\frac{dA_2^{2\omega}}{dy} = \frac{i\omega}{Neff_2^{2\omega}c}\left[\kappa_2^2 (A_0^\omega)^2 \cdot e^{i\Delta\beta_2 y}\right], \quad (3)$$

where $A_n^{\omega(2\omega)}$ represents the amplitude of the $n^{th}$ order of FW (SH) mode; $Neff_n^{\omega(2\omega)}$ represents the effective refractive index of the $n^{th}$ order of FW (SH) mode; $\omega$ is the circular frequency of FW; $\kappa_1^{0(2)}$ and $\kappa_2^{0(2)}$ are nonlinear coupling coefficients and can be expressed as:

$$\kappa_1^{0(2)} = \frac{\int\left(\overrightarrow{E_0^\omega}(z)\right)^* \cdot \chi^{(2)} : \overrightarrow{E_{0(2)}^{2\omega}}(z)\left(\overrightarrow{E_0^\omega}(z)\right)^* dz}{\int \overrightarrow{E_0^\omega}(z) \cdot \left(\overrightarrow{E_0^\omega}(z)\right)^* dz} \quad (4)$$

$$\kappa_2^{0(2)} = \frac{\int\left(\overrightarrow{E_{0(2)}^{2\omega}}(z)\right)^* \cdot \chi^{(2)} : \overrightarrow{E_0^\omega}(z)\overrightarrow{E_0^\omega}(z) dz}{\int \overrightarrow{E_{0(2)}^{2\omega}}(z) \cdot \left(\overrightarrow{E_{0(2)}^{2\omega}}(z)\right)^* dz}, \quad (5)$$

in which $E_n^{\omega(2\omega)}$ represents the transverse field distribution of the $n^{th}$ order of FW (SH) mode and $\chi^{(2)}$ is nonlinear susceptibility tensor; $\Delta\beta_j$ is the momentum mismatch of the nonlinear interaction between fundamental mode of FW and the $j^{th}$ mode of SH. In addition, as discussed above, mode conversion process induced by PT symmetry should be coupled with these nonlinear interactions simultaneously. Thus, another two dynamic equations should be included:

$$\frac{dA_0^{2\omega}}{dy} = i\gamma_1 \xi(y) A_0^{2\omega}(y) + i\gamma_2 \xi(y) A_2^{2\omega}(y) \cdot e^{-i\Delta\beta_3 y} \tag{6}$$

$$\frac{dA_2^{2\omega}}{dy} = i\gamma_3 \xi(y) A_0^{2\omega}(y) e^{i\Delta\beta_3 y} + i\gamma_4 \xi(y) A_2^{2\omega}(y), \tag{7}$$

in which the coupling coefficients are described as follows:

$$\gamma_1 = \frac{\omega^2}{2(\beta_0^{2\omega}) \cdot c^2} \frac{\int \left(\overrightarrow{E_0^{2\omega}}(z)\right)^* \Delta\xi \overrightarrow{E_0^{2\omega}}(z) dz}{\int \left|\overrightarrow{E_0^{2\omega}}(z)\right|^2 dz} \tag{8}$$

$$\gamma_2 = \frac{\omega^2}{2(\beta_0^{2\omega}) \cdot c^2} \frac{\int \left(\overrightarrow{E_0^{2\omega}}(z)\right)^* \Delta\xi \overrightarrow{E_2^{2\omega}}(z) dz}{\int \left|\overrightarrow{E_0^{2\omega}}(z)\right|^2 dz} \tag{9}$$

$$\gamma_3 = \frac{\omega^2}{2(\beta_2^{2\omega}) \cdot c^2} \frac{\int \left(\overrightarrow{E_2^{2\omega}}(z)\right)^* \Delta\xi \overrightarrow{E_0^{2\omega}}(z) dz}{\int \left|\overrightarrow{E_2^{2\omega}}(z)\right|^2 dz} \tag{10}$$

$$\gamma_4 = \frac{\omega^2}{2(\beta_2^{2\omega}) \cdot c^2} \frac{\int \left(\overrightarrow{E_2^{2\omega}}(z)\right)^* \Delta\xi \overrightarrow{E_2^{2\omega}}(z) dz}{\int \left|\overrightarrow{E_2^{2\omega}}(z)\right|^2 dz}, \tag{11}$$

$\Delta\beta_3 = \beta_0^{2\omega} - \beta_2^{2\omega}$, and $\Delta\xi$ represents the modulation depth of the perturbation function.

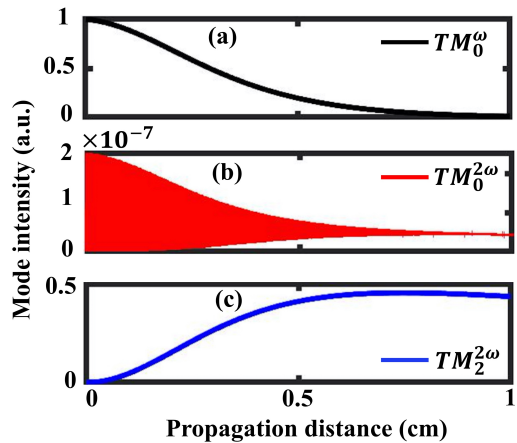

Fig. 2. The dependences of nonlinear mode conversion on propagation distance without PT-symmetry modulation. (a) $TM_0^\omega$ mode of FW. (b) $TM_0^{2\omega}$ mode of SH. (c) $TM_2^{2\omega}$ mode of SH.

If no PT structure is loaded, i.e. $\Delta\xi = 0$, nonlinear interaction will play a dominant role in the propagation. In general, frequency-doubling efficiency of each SH mode is decided collaboratively by both nonlinear coupling coefficient (overlapping integral of guided modes) and phase-matching condition. Therefore, in our simulations, we take into account two cases of nonlinear transfer of $TM_2^{2\omega}$ and $TM_0^{2\omega}$ modes. As depicted in Fig. 2, with the depletion of FW, the weak intensity of $TM_0^{2\omega}$ mode oscillates with propagation distance and ultimately tend to a stable value. While for $TM_2^{2\omega}$ mode, its intensity, just as we designed, grows with the increase of length and is about six orders of magnitude higher than that of $TM_0^{2\omega}$ mode. This result also indicates that phase-matching is the key factor, although $\kappa_1^0$, $\kappa_2^0$ for $TM_0^{2\omega}$ mode are about 23 times and 2 times $\kappa_1^2$, $\kappa_2^2$ for $TM_2^{2\omega}$ mode based on the calculations in our model respectively. Due to the reciprocity of frequency-doubling, the simulation results are identical, either for forward-propagating or backward-propagating of FW.

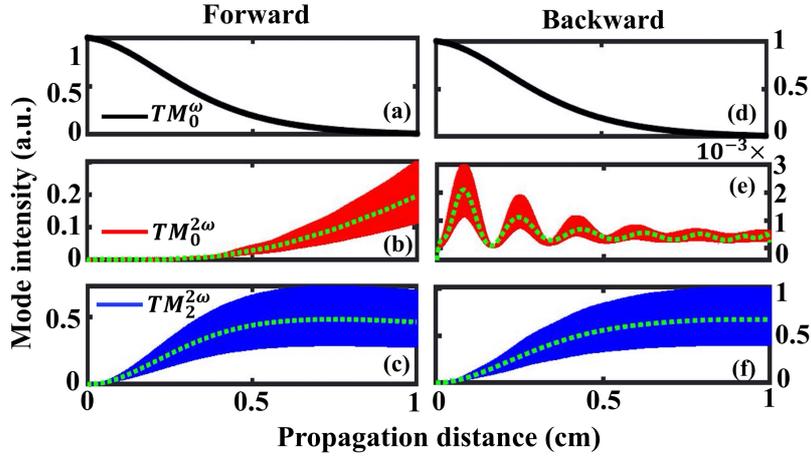

Fig. 3. Asymmetric mode conversion with nonlinearity and PT-symmetry modulation together. The mode intensities, respectively for $TM_0^\omega$ mode of FW, $TM_0^{2\omega}$ mode of SH and $TM_2^{2\omega}$ mode of SH, vary with the change of propagation distance under the situations of forward-propagation (a)-(c) and backward-propagation (d)-(f).

Subsequently, we consider the situations with PT modulation. Then, for the forward case, Eqs. (6) and (7) are revised as,

$$\frac{dA_0^{2\omega}}{dy} = i\gamma_1 A_0^{2\omega}(y)e^{iqy} + i\gamma_2 A_2^{2\omega}(y) \cdot e^{-i(\Delta\beta_3 - q)y} \qquad (12)$$

$$\frac{dA_2^{2\omega}}{dy} = i\gamma_3 A_0^{2\omega}(y)e^{i(q+\Delta\beta_3)y} + i\gamma_4 A_2^{2\omega}(y)e^{iqy}. \tag{13}$$

In terms of our calculated results of $\Delta\beta_3$, the selected $q$ is set to be 1.4226 $\mu m^{-1}$ under which the second exponential term of Eq. (12) will vanish. Meanwhile, these two procedures, nonlinear SH and mode conversion induced by PT, are tightly coupled with each other, and the nonlinear process provides an essential value of $A_2^{2\omega}$. Consequently, the intensity of $TM_0^{2\omega}$ mode is enhanced as the propagation distance increases, as shown in Fig. 3(b). On the other hand, in virtue of the one-way nature of $\vec{q}$, both the exponential terms of Eq. (13) cannot be neglected, which limits the chance that the generated $TM_0^{2\omega}$ mode turns back into $TM_2^{2\omega}$ mode. So, the contribution to $TM_2^{2\omega}$ mode intensity mainly comes from nonlinear SH interaction. As demonstrated in Fig. 3(a) and 3(c), the intensity of $TM_0^{\omega}$ mode is gradually converted into that of $TM_2^{2\omega}$ mode owing to the phase matching between these two modes. Besides, it is seen $TM_2^{2\omega}$ mode power remains oscillation as it grows with the propagation length. This is also caused by the coupling between the aforesaid two primary processes. Hence, when $TM_2^{2\omega}$ mode propagates along the *y* direction, the phase-mismatch enlarged by $\vec{q}$ in Eq. (13) will result in such oscillation along with a phase-matched SH process. Moreover, as illustrated in Fig. 3(b), this oscillation feature also takes place for $TM_0^{2\omega}$ mode, which is principally determined by two aspects consisting of the oscillating $TM_2^{2\omega}$ mode intensity and the residual exponential term in Eq. (12). To reveal clearly these features, the polynomial fitting curves (green dot lines) have been calculated and displayed in Fig.3.

While for the backward case, the dynamics can be derived as:

$$\frac{dA_0^{2\omega}}{dy} = -i\gamma_1 A_0^{2\omega}(y)e^{iqy} - i\gamma_2 A_2^{2\omega}(y) \cdot e^{i(q+\Delta\beta_3)y} \tag{14}$$

$$\frac{dA_2^{2\omega}}{dy} = -i\gamma_3 A_0^{2\omega}(y)e^{i(q-\Delta\beta_3)y} - i\gamma_4 A_2^{2\omega}(y)e^{iqy}. \tag{15}$$

In contrast with the forward case, the selected $q$ cannot compensate for the mismatch in Eq. (14), but magnifies it instead. It somewhat restrains $TM_2^{2\omega}$ mode from converting into $TM_0^{2\omega}$ mode. As depicted in Fig. 3(e), the intensity of $TM_0^{2\omega}$ mode gradually oscillates to a nearly stable value which is approximately three orders lower than that in the forward case. In addition, one can find the first exponential term of Eq. (15) can be cancelled, indicating that $TM_0^{2\omega}$ mode could be transferred into $TM_2^{2\omega}$ mode. However, in our scheme, the conversion from $TM_0^{\omega}$ mode to $TM_0^{2\omega}$ mode, as demonstrated in Fig. 2(b), is extremely inefficient, which means nonlinear interaction cannot offer sufficient $A_0^{2\omega}$ to support this PT-symmetry induced interaction. Hence for $TM_2^{2\omega}$ mode as shown in Fig. 3(f), it is analogous to the forward case that the intensity is decided largely in frequency-doubling process.

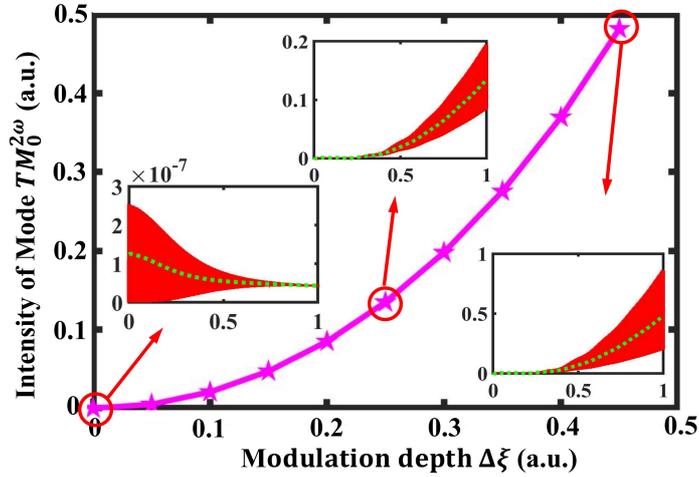

Fig. 4 The excited $TM_0^{2\omega}$ mode intensity changes with the variation of the modulation depth of PT-symmetry. The insets illustrate the mode intensities that can be eventually generated with distinct modulation depth $\Delta\xi = 0$, $\Delta\xi = 0.25$ and $\Delta\xi = 0.45$.

The foregoing comparative analysis of Fig. 3 demonstrates such a mode conversion in waveguide, with the coupling of nonlinearity and PT symmetry, can be realized and asymmetric. Owing to the involvement of PT symmetry, this type of mode conversion correlates strongly with the modulation depth $\Delta\xi$. As illustrated in Eqs. (6)-(7), the coupling coefficients $\gamma_i$ that are attributable to PT symmetry are proportional to $\Delta\xi$. For the forward propagation, the output intensity of $TM_0^{2\omega}$ mode mainly relies on the coefficient $\gamma_2$. As shown in Fig. 4, it is found that a greater modulation depth will achieve a higher value of $\gamma_2$, and consequently leads to a larger transformed mode intensity that grows exponentially with $\Delta\xi$. Besides, the other three coefficients $\gamma_1$, $\gamma_3$ and $\gamma_4$ determine the oscillation property of output modes. From the insets of Fig. 4, respectively for $\Delta\xi = 0$, $\Delta\xi = 0.25$ and $\Delta\xi = 0.45$, we can clearly see that the amplitude of oscillation rises with the increase of $\Delta\xi$.

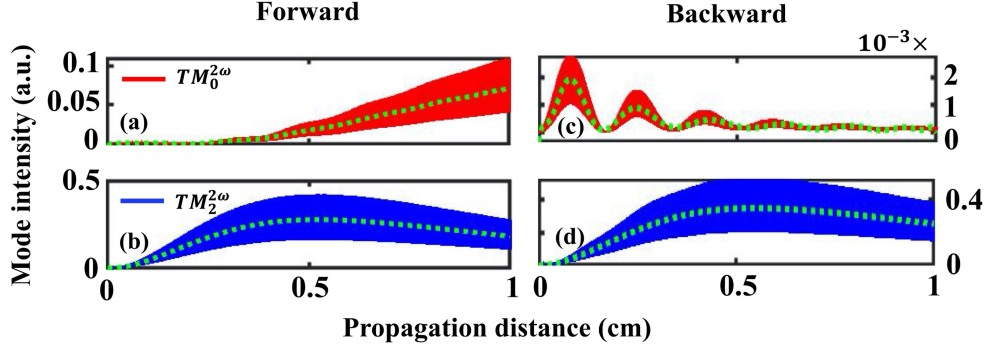

Fig. 5 Asymmetric NMC in a purely passive system respectively for the forward (a)-(b) and backward (c)-(d) propagations.

For the above theoretical analysis, we employ the ideal model of PT symmetry with balanced gain and loss for the one-way wave-vector to demonstrate asymmetric NMC, while similar results could be realized by considering other simple ways, such as passive PT systems [19, 24]. In terms of our simulations where a purely passive system is taken, the performance of NMC is demonstrated in Fig. 5. Compared with the ideal case, the efficiency of NMC shown in Fig. 5(a) becomes lower. On the other hand, despite no linear gain, nonlinear SH process provides an initial stimulation of the cascaded mode conversion induced by PT, when the pump propagates over each short distance in the waveguide. Meanwhile, the efficiency of this nonlinear process is proportional to the square of propagation length, which implies that the stimulation to NMC will enhance along the propagation direction of pump. Consequently, the efficiency of NMC can continue to grow provided that the nonlinear process is not particularly inefficient. This feature is significantly distinct from that in the case of a single linear passive system without nonlinearity. In that case, the stimulation to mode conversion gradually diminishes with the growing distance, which will make the conversion efficiency increase first and then decrease. Obviously, the reason leading to this difference in efficiency, in contrast with our NMC scheme, lies in the coupling between the nonlinear and linear interactions.

For the discussions above, we are primarily concerned with the NMC in the identical direction, i.e. the converted SH propagates along the input direction of FW. However, utilizing our scheme, can 'reflected' NMC be implemented? For instance, the system is excited by a negative-*y*-propagating FW mode (the backward case in Fig. 1(a)), generating a positive-*y*-propagating ('reflected') SH mode. In traditional nonlinear optics, such a counter propagating coupling, in general, is tough to be achieved, since the 'reflected' SH and the FW modes cannot fulfill the conservation of momentum. To overcome this issue, the most direct way is to address nonlinear process itself, such as employing negative-index [37] or zero-index [38, 39] metamaterials. Unlike this kind of thought, we can transfer the momentum (phase) mismatch from nonlinear process to linear process, owing to the coupling between the nonlinear and linear interactions assisted by PT symmetric modulation. Therefore, in our case, the negative-*y*-propagating FW mode will still effectively generate negative-*y*-propagating SH mode $TM_2^{2\omega}$ with a propagation constant $\overrightarrow{\beta_2^{2\omega}}$, which is dominated by the mode dispersion in Fig. 1(b). And then, this negative-*y*-propagating SH mode is subsequently converted into the 'reflected' SH mode $TM_2^{2\omega}$ with a propagation constant $-\overrightarrow{\beta_2^{2\omega}}$, provided that $\overrightarrow{\beta_2^{2\omega}} - \vec{q} = -\overrightarrow{\beta_2^{2\omega}}$. This relationship is different from that in the case of Fig. 3 or Fig. 5 where $\vec{q}$ is so small that $\overrightarrow{\beta_2^{2\omega}} - \vec{q}$ is located in the non-guided region of Fig. 1(c). Therefore, we only require a lager $\vec{q} = 2\overrightarrow{\beta_2^{2\omega}}$ that can make $\overrightarrow{\beta_2^{2\omega}}$ 'flip' over the non-guided region to the positive-*y* region, and thus realizing 'reflected' NMC. Under this circumstance, if the system is excited by a positive-*y*-propagating FW mode (the forward case in Fig. 1(a)), the phase mismatch for counter propagating coupling still exists so that 'reflected' NMC cannot take place. Consequently, it is analogous to the case in Fig. 3 that such type of 'reflected' NMC remains asymmetric.

## 4. Conclusion

In conclusion, we have demonstrated the asymmetric NMC by combining frequency-doubling process and periodic PT symmetric modulation in a planar LN waveguide. Following the basic principle, asymmetric generation of SH modes from FW could be arbitrarily designed. For example, this scheme of asymmetric NMC can be potentially applied in a multi-channel configuration, in which each channel is engineered with a reasonable modulation period, and then asymmetric switchover between several desired SH modes on a single chip could be manipulated in a diverse way. Note that we only consider TM mode conversion coupled with nonlinear interaction type of $TM + TM \Rightarrow TM$ for above discussion. Actually, there are other three interaction options in general: $TM + TM \Rightarrow TE$, $TE + TE \Rightarrow TE$ and $TE + TE \Rightarrow TM$, and effective asymmetric NMC is greatly dependent on the efficiency of nonlinear process. However, the nonlinear coupling coefficients or dominated nonlinear coefficients ($d_{22}$ and $d_{31}$) in these three cases are pretty small, so as to result in inefficient nonlinear mode couplings and interactions. Meanwhile, the frequency-doubling process here could be extended to other nonlinear interactions, including sum-frequency generation, third harmonic generation and four-wave mixing. Our proposed way, therefore, could be further developed to manipulate nonlinear fields with asymmetric response, which can provide a feasible avenue toward designing new nonlinear devices enabled by non-Hermitian optics and meta-optics.

**Funding.** National Natural Science Foundation of China (11904169, 91950106); Natural Science Foundation of Jiangsu Province (BK20190383); Postdoctoral Science Foundation of Jiangsu Province (2020Z224); China Postdoctoral Science Foundation (2020M681576).

**Acknowledgment.** This work is supported by Top-notch Academic Programs Project of Jiangsu Higher Education Institutions.

**Disclosures.** The authors declare no conflicts of interest.

**Data Availability.** Data underlying the results presented in this paper are not publicly available at this time but may be obtained from the authors upon reasonable request.

**References:**

1. C. M. Bender and S. Boettcher, Phys. Rev. Lett. **80**, 5243-5246 (1998).

2. G. Y. Sun, J. C. Tang, and S. P. Kou, Front. Phys. **17**, 33502 (2022).

3. R. E. Ganainy, K. G. Makris, M. Khajavikhan, Z. H. Musslimani, S. Rotter and D. N. Christodoulides, Nat. Phys. **14**, 11-19 (2017).

4. M. A. Miri and A. Alu, Science **363**, eaar7709 (2019).

5. A. Guo, G. J. Salamo, D. Duchesne, R. Morandotti, M. V. Ravat and V. Aimez, Phys. Rev. Lett. **103**, 093902 (2009).

6. Y. Fu, Y. Xu and H. Chen, Opt. Express **24**,1648(2016).

7. A. Laha, S. Dey, H. K. Gandhi, A. Biswas and S. Ghosh, ACS Photonics **7**, 967–974 (2020).

8. H. Hodaei, M. A. Miri, M. Heinrich, D. N. Christodoulides and M. Khajavikhan, Science **346**, 975-978 (2014).

9. B. Peng, S. K. Ozdemir, M. Liertzer, W. J. Chen, J. Kramer, H. Yilmaz, J. Wiersig, S. Rotter and L. Yang, PNAS **113**, 6845-6850 (2016).

10. N. Zhang, Z. Y. Gu, K. Y. Wang, M. Li, L. Ge, S. M. Xiao and Q. H. Song, *Laser Photon. Rev.* **11,** *1700052 (2017).*

11. L. Chang, X. S. Jiang, S. Y. Hua, C. Yang, J. M. Wen, L. Jiang, G. Y. Li, G. Z. Wang and M. Xiao, Nat. Photon. **8**, 524-529 (2014).

12. B. Peng, S. K. Ozdemir, F. C. Lei, F. Monifi, M. Gianfreda, G. L. Long, S. H. Fan, F. Nori, C. M. Bender and L. Yang, Nat. Phys. **10**, 394-398 (2014).

13. C. D. Chen and L. N. Zhao, Opt. Commun. **474**, 126108 (2020).

14. C. D. Chen, Y. J. Xie and S. W. Huang, Opt. Commun. **483**, 126674 (2021).

15. M. Chen, Z. F. Li, X. Tong, X. D. Wang, and F. H. Yang, Opt. Express **28**, 17966 (2020).

16. Y. Liang, Q. Gaimard, V. Klimov, A. Uskov, H. Benisty, A. Ramdane, and A. Lupu, Phys. Rev. B **103**, 045419 (2021).

17. Y. Cao, Y. Fu, Q. Zhou, Y. Xu, L. Gao, and H. Chen, Opt. Express **27**,7857(2019).

18. Y. Fu, Y. Fei, D. Dong, and Y. Liu, Front. Phys. **14**, 62601(2019).

19. L. Feng, R. El-Ganainy, and L. Ge, Nat. Photon. **11**,752-762(2017).

20. M. Greenberg and M. Orenstein, Opt. Lett. **29**, 451-453 (2004).

21. L. Feng, M. Ayache, J. Q. Huang, Y. L. Xu, M. H. Lu, Y. F. Chen, Y. Fainman and A. Scherer, Science **333**, 729-733 (2011).

22. S. N. Ghosh and Y. D. Chong, *Sci. Rep.* **6,** 19837 (2016).

23. D. Chatzidimitriou, A. Pitilakis, T. Yioultsis, and E. E. Kriezis, Phys. Rev. A **103**, 053503 (2021).

24. A. Pitilakis, D. Chatzidimitriou, T. V. Yioultsis, and E. E. Kriezis, IEEE J. Quant. Electron. **57**, 8400210 (2021).

25. Z. Lin, H. Ramezani, T. Eichelkraut, T. Kottos, H. Cao and D. N. Christodoulides, Phys. Rev. Lett. **106**, 213901 (2011).

26. L. Feng, Y. L. Xu, W. S. Fegadolli, M. H. Lu, J. E. B. Oliveira, V. R. Almeida, Y. F. Chen and A. Scherer, Nat. Mater. **12**, 108–113 (2013).

27. Y. F. Jia, Y. X. Yan, S. V. Kesava, E. D. Gomez and N. C. Giebink, ACS Photonics **2**, 319−325 (2015).

28. X. Y. Zhu, Y. L. Xu, Y. Zou, X. C. Sun, C. He, M. H. Lu, X, P. Liu and Y. F. Chen, Appl. Phys. Lett. **109**, 111101 (2016).

29. H. Zhao, W. S. Fegadolli, J. K. Yu, Z. F. Zhang, L. Ge, A. Scherer and L. Feng, Phys. Rev. Lett. **117**, 193901 (2016).

30. W. Wang, L. Q. Wang, R. D. Xue, H. L. Chen, R. P. Guo, Y. M. Liu and J. Chen, Phys. Rev. Lett. **119**, 077401 (2017).

31. Y. Y. Fu, Y. D. Xu, and H. Y. Chen, Front. Phys. **13**, 134206 (2018).

32. L. Chang, Y. F. Li, N. Volet, L. Wang, J. Peters and J. E. Bowers, Optica **3**, 531-535 (2016).

33. M. Jankowski, C. Langrock, B. Desiatov, A. Marandi, C. Wang, M. Zhang, C. R. Phillips, M. Loncar and M. M. Fejer, Optica **7**, 40-46 (2020).

34. D. H. Sun, Y. W. Zhang, D. Z. Wang, W. Song, X. Y. Liu, J. B. Pang, D. Q. Geng, Y. H. Sang and H. Liu, Light: Science & Applications **9**, 197 (2020).

35. R. Luo, Y. He, H. X. Liang, M. X. Li and Q. Lin, Optica **5**, 1006-1011 (2018).

36. C. Wang, Z. Y. Li, M. H. Kim, X. Xiong, X. F. Ren, G. C. Guo, N. F. Yu and M. Loncar, Nat. Commun. **8**, 2098 (2017).

37. A. Rose, D. Huang, and D. R. Smith, Phys. Rev. Lett. **107**, 063902 (2011).

38. A. Rose, and D. R. Smith, Opt. Mater. Express **1**, 1232-1243 (2011).

39. H. Suchowski, K. O'Brien, Z. J. Wong, A. Salandrino, X. Yin, and X. Zhang, Science **342**, 1223 (2013)